# Passenger's Behavior in Response to an Unplanned Transit Disruption

Nima Golshani [a], Ehsan Rahimi [a], Ramin Shabanpour [a], Kouros Mohammadian [a],
Joshua Auld [b], Hubert Ley [c]

[a] Department of Civil & Materials Engineeing, University of Illinois at Chicago
[b] Systems Modelling and Controls Group, Argonne National Laboratory
[c] Transportation Research and Analysis Computing Center, Argonne National Laboratory

April 2019

**ABSTRACT**

Public transit disruption is becoming more common across different transit services, which can have a destructive influence on the resiliency and reliability of the transportation system. Even though transit agencies have various strategies to mitigate the probability of failure in the transit system by conducting preventative actions, some disruptions cannot be avoided due to their either unpredictable or uncontrollable nature. Utilizing a recently collected data of transit users in the Chicago Metropolitan Area, the current study aims to analyze how transit users respond to an unplanned service disruption and disclose the factors that affect their behavior. In this study, a random parameter multinomial logit model is employed to consider heterogeneity across observations as well as panel effects. The results of the analysis reveal that a wide range of factors including socio-demographic attributes, personal attitudes, trip-related information, and built-environment are significant in passengers' behavior in case of unplanned transit disruptions. Moreover, the effect of service recovery time on passengers are not the same among all types of disrupted services; rail users are more sensitive to the recovery time as compared to bus users. The findings of this study provide insights for transportation authorities to improve the transit service quality in terms of user's satisfaction and transportation resilience. These insights help transit agencies in order to implement effective recovery strategies.

*Keywords:* public transit, unplanned disruption, SP-RP survey, recovery time, travel behavior

## INTRODUCTION

With a population of 9.5 million people, Chicago metropolitan area is the third largest metropolitan area in the United States which has one of the highest density of people and businesses in the country ("Census Reporter," 2017). The region, however, is vulnerable to different types of hazards including tornadoes, blizzards and man-made disasters (Eitel, 2018). In such emergency conditions, transportation system resilience, and more specifically transit system resilience, can help the affected areas to effectively respond to and recover from such situations (Auld et al., 2018). The Chicago Transit Authority (CTA) services over 3.5 million riders in the city of Chicago and 35 suburbs surrounding the city (Chicago Transit Authority, 2017). CTA provides the nation's second-largest public transportation system which operates under the budget of over $1.5k million.

Public transit disruption is becoming more common across different transit services which can have a destructive influence on the resiliency and reliability of the transportation system. Even though transit agencies including CTA have various strategies to mitigate the probability of failure in the transit system by conducting preventative actions, some disruptions cannot be avoided due to their either unpredictable or uncontrollable nature. Service disruptions are typically classified into two categories of pre-planned and unplanned events. A pre-planned disruption occurs as a result of disruptive activities planned ahead of time (Pnevmatikou et al., 2015; van Exel and Rietveld, 2009; Yap et al., 2018). Examples of such disruptions include road or rail closures due to maintenance and labor strikes. On the other hand, unplanned disruptions are mostly due to uncontrolled incidents such as earthquakes, floods, and storms (as natural disasters) or terrorist attacks (as man-made disasters). This scope of this study is focused on the unplanned disruptions.

During an unplanned service disruption, the top priority for passengers (i.e., reaching their destinations in a timely manner) does not necessarily align with the top priority for the transit authorities (i.e., mitigating the impact of disrupted service in terms of time and distance) (Lin, 2017). Thus, understanding transit users' behavior in response to such conditions can assist those authorities to plan for more effective recovery strategies. However, transit users' behavior during an unplanned service disruption in a multimodal transportation network has not been fully investigated yet (Papangelis et al., 2016; Zhu et al., 2017). To fill such an important gap, the present study looks deep into diverse aspects of transit users' response behavior to understand how an unplanned disruption in the transit system affects their travel preferences. In case of facing a disrupted service, one may cancel the trip, change the destination, or switch to other travel modes such as personal vehicle, taxi, or ride-sharing services. Therefore, to gain a comprehensive understanding of people's decision behavior in such cases, it is vital to consider all these potential alternatives.

Private vehicles are typically the biggest competitor to transit service. Service reliability, privacy, convenience, availability, and travel time are among the significant dis-utilities of transit services compared to private vehicles. Further, the emergence of new mobility options such as ride-hailing services has changed the game in the transportation market. Today, not only does the transportation network companies (TNCs) take market share from private transport options, but also, they are strong competitors of the transit system. These companies provide a wide variety of affordable, door-to-door options and, thereby, encourage transit users to substitute their conventional travel choices with these flexible services.

Based on an intercept stated preference (SP)-revealed preference (RP) survey, which is recently conducted in the Chicago metropolitan area, the current research is designed to scrutinize Chicago transit riders' decision behavior in case of facing an unplanned service disruption. In a specific part of the survey, respondents were presented with multiple hypothetical disruption scenarios and were asked to indicate which action they would likely take while the intercepted transit service is disrupted. In each scenario, multiple options are offered including canceling the trip, changing the destination, waiting for a back-up shuttle bus, or switching to other modes such as asking a family member for a ride, picking up personal vehicle, taking a taxi, or using a ride-sharing service. The collected data is enriched by adding several built-environment indicators including population density, land-use mix, neighborhood design, pedestrian-oriented network density, destination accessibility, transit accessibility and frequency, among others. Finally, a random parameter multinomial logit model is estimated to analyze respondents' decision behavior in this study.

Rest of this paper is organized as follows. The next section provides a brief review of related studies on transit users' behavior during service disruptions. Then, the conducted SP-RP survey and incorporated data from other data sources are elaborated. Following that, an overview of the applied model and its estimation process is briefly presented. Finally, the findings of this study are elaborated in the results section.

**LITERATURE REVIEW**
There is a limited but growing body of literature on transit users' behavior in case of service disruptions. These studies employ various types of data sources including RP, SP, and SP-RP surveys and apply different methodological approaches. In this section, we briefly review some of the most recent published works in the area.

As one of the first studies in this area, Murray-Tuite et al. (2014) investigated the long-term impacts of the deadly Metrorail collision (that took place in June 2009 in Washington, D.C.) on passengers' travel behavior. Using a web-based RP survey, respondents, who had used the Metrorail during the six months before the incident, were asked to specify what changes they made to their transit trips in terms of mode and seat location after the collision. The researchers found out that about 10% and 17% of respondents had altered their travel mode and seating location in the same train, respectively.

Due to limitations of RP surveys such as insufficient variation in the data to investigate all variables of interest and potential strong correlations between explanatory variables (Kroes and Sheldon, 1988), some scholars suggested use of stated-preference (SP) surveys to reveal transit users' behavior during service disruptions. In an SP survey, respondents are presented with one or multiple hypothetical scenarios and are asked to indicate their decisions when facing such a situation in the real world. Conducting an SP survey from train passengers in Klang Valley, Malaysia, Bachok (2008) focused on modal shift behavior of rail users. In this study, train passengers were asked to choose an alternative among a set of provided options, including other trains, shuttle bus, private vehicles, and wait for the restoration of the rail system in a hypothetical scenario.

Similarly, Fukasawa et al. (2012) investigated the effect of providing information such as estimated arrival time, arrival order and congestion level on passengers' modal shift behavior in response to unplanned transit disruption using a data from an SP survey. They found that train users, who have access to the information, generally have a higher tendency of shifting to other trains in comparison with those without access to the information. In contrast, Bai and Kattan (2014) conducted an SP survey on light rail transit passengers in Calgary, Canada, and found out that respondents without access to the information concerning possible recovery period have more willingness to switch their travel mode.

Since SP surveys may not necessarily represent transit users' behavior in a real service disruption incident (Kroes and Sheldon, 1988), more recent studies have suggested combining both SP and RP surveying methods. For instance, Lin et al. conducted a combined RP-SP survey in Toronto area to analyze transit users' mode choice behavior in response to a subway disruption (Lin, 2017; Lin et al., 2016). The RP section of the survey was devoted to respondents' last experience with an unplanned service disruption and the SP section provided hypothetical disruption scenarios in which respondents were asked to either choose among alternative modes or cancel their trip. They found that travel cost, waiting time, duration of delay, income, and type of incident could affect transit users' commuting mode choice during a subway service disruption (Lin et al., 2016).

As an alternative to the recall SP-RP survey conducted by Lin et al. (2016), we have designed a new SP-RP survey framework to intercept the respondents while waiting in the transit stations. Intercept surveys reduce memory effect, mainly if they are conducted during or immediately after a trip (Sudman and Bradburn, 1973). In addition, we have attempted to enrich our data by collecting information about riders' personal attitudes and built-environment factors. Further information about the survey design and collected data is provided in the following section.

## DATA

The main source of data used in the current study is a survey recently designed and conducted by a research team from Argonne National Laboratory, University of Illinois at Chicago, and University of Chicago. Full information about design of various parts of the survey, implementation process, and summary statistics of the collected data can be found in authors' previous work (Auld et al., 2018). Here, we only elaborate on the sections that are related to the scope of the current study.

The survey was designed as an in-station intercept survey with the objective of analyzing transit riders' decision behavior in case of facing an unplanned service disruption in the Chicago metropolitan area. Passengers were intercepted by the survey implementation team at about 100 separate stops/stations belonging to all major operators of transit network in the region including PACE (suburban bus operator), Metra (commuter rail) and CTA (urban bus & rail operator). In order to achieve a sample which is a good representative of the population of interest, a sampling plan was developed considering average daily ridership as well as boarding information of each agency.

The survey was designed using a web-based surveying platform and it was accessible through a survey link and PIN distributed to the intercepted respondents. Passengers who agreed to participate in the study were given a contact card with a unique PIN which identified the service, contact time, and station. Through entering the PIN, respondents were directed to the online questionnaire. In the survey, a full set of individual- and household-level socio-demographic information as well as comprehensive information about the intercepted transit trip including fare, time, origin, destination, access/egress, ride quality, time use, and trip purpose among others were collected. Further, passengers' preferences and attitudes towards transit and other mobility options such as taxi, ride-sharing, and bike-sharing programs were collected. With respect of the scope of the current study, respondents were presented with multiple hypothetical disruption scenarios based on the intercepted transit trip and were asked to indicate which action they would likely take while the intercepted transit service is disrupted (**Figure 1**). Their response to this question has been considered as the variable of interest (dependent variable) in the current study.

In each scenario, seven potential actions were listed including waiting for a back-up shuttle bus, asking for a ride from family/friend, picking up his/her own auto (if any), taking a taxi, using ride-sharing services, changing the trip destination, and canceling the trip. Each option, if applicable, was further described in terms of waiting time, travel time, arrival time, and cost.

Alternative-specific attributes were generated considering the characteristics of the intercepted trip as the basis for a set of SP questionnaires with randomly altered modal characteristics set according to an experimental design which is fully described in Auld et al. (2018). The information displayed to the respondents all pivots off of the exact transit and driving trip characteristics as determined by the Google Direction API router at the actual time of departure, so real-time traffic congestion, transit schedule, etc. are accounted for when setting the scenario values. Further, a set of piece-wise functions was used to generate travel cost and waiting times. **Figure 1** illustrates an example of an SP transit disruption scenario presented to each respondent.

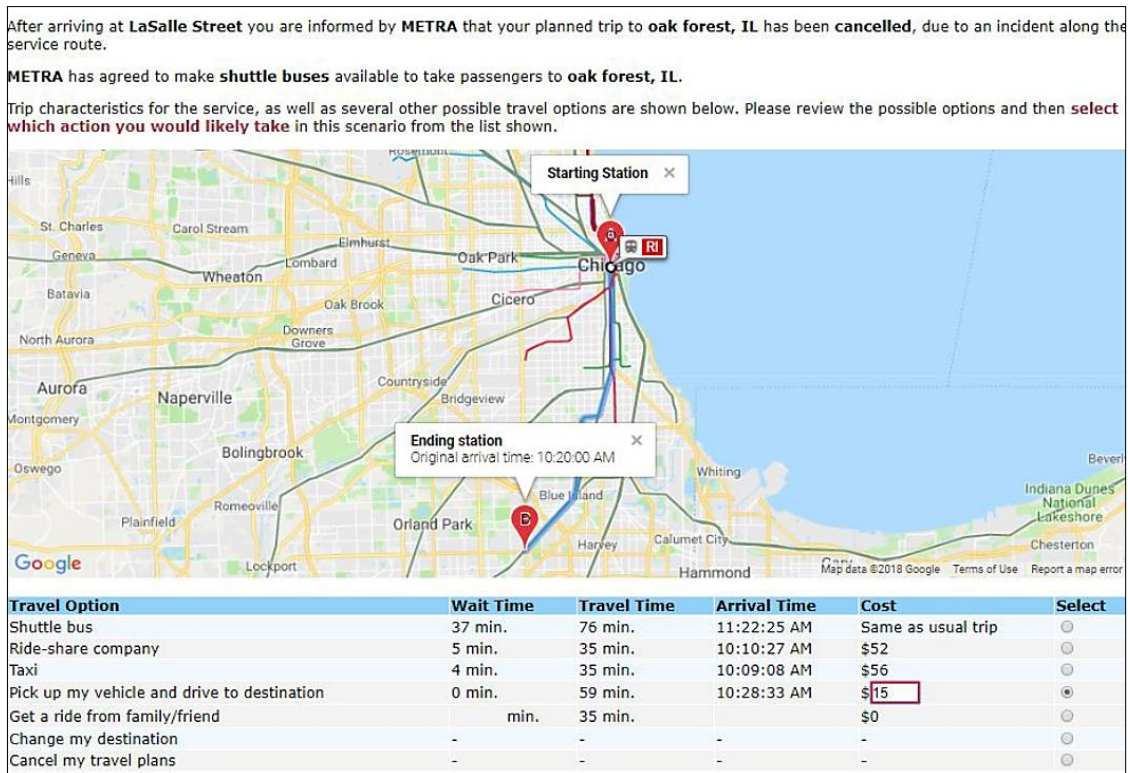

**Figure 1.** An example of an SP transit disruption scenario (Adapted from Auld et al. (2018)).

After rejecting observations with either missing or invalid information, the final database for this study includes 628 respondents who were faced with 2498 disruption choice situations. The collected dataset consists of 45.2% male and 54.0% female participants who live in Chicago metropolitan area. As for the age, 17.6% are less than 24 years old, 32.8% are between 25 and 34, 19.6% are between 35 and 44, and the rest are older than 45 years old. As for the employment status, the data contains 72.1% full-time workers, 10.6% part-time workers, and the rest consists of unemployed and retired people. **Table 1** presents summary statistics of respondents' key demographic attributes in the collected sample.

**TABLE 1**  Summary Statistics Of The Respondents' Key Characteristics.

| Variable | Category | Share (percentage) |
|---|---|---|
| Household size | 1 | 28.53 |
| | 2 | 47.80 |
| | 3 | 13.66 |
| | 4 | 5.92 |
| | 5 or more | 3.49 |
| Household income | Under $15K | 6.37 |
| | $15K - $35K | 8.95 |
| | $35K - $50K | 11.99 |
| | $50K - $75K | 12.90 |
| | $75K - $100K | 14.72 |
| | $100K or more | 30.80 |
| Gender | Male | 45.22 |
| | Female | 54.02 |
| Age | ≤ 24 | 17.60 |
| | 25-34 | 32.78 |
| | 35-44 | 19.58 |
| | 45-54 | 15.02 |
| | 55-64 | 11.99 |
| | ≥ 65 | 2.89 |
| Race | White/Caucasian | 56.90 |
| | African American | 16.54 |
| | Hispanic/Latino | 10.47 |
| | Asian | 8.50 |
| | Two or more ethnicities | 4.10 |
| | Native American | 0.61 |
| | Other | 1.97 |
| Education | Less than high school | 0.91 |
| | High school graduate | 5.31 |
| | Some college credit | 13.96 |
| | Vocational school certificate | 1.06 |
| | Associate degree | 6.53 |
| | Bachelor's degree | 37.94 |
| | Graduate degree | 33.38 |
| Employment status | Full time | 72.08 |
| | Part time | 10.62 |
| | Other | 17.30 |

Note: The sum of the percentages may not equal 100 due to observations with missing values

It should be noted that the personal vehicle option was not available for those who either had no vehicle or had no access to their vehicle at the time of the disruption. **Figure 2** presents the distribution of selected actions with respect to the personal vehicle availability. As can be seen in the figure, waiting for the back-up shuttle and using ride-sharing services are the first and second most frequent selected actions in the data. It is also interesting to note that in about 10% and 5% of scenarios, people decided to cancel their trip or change their destination, respectively.

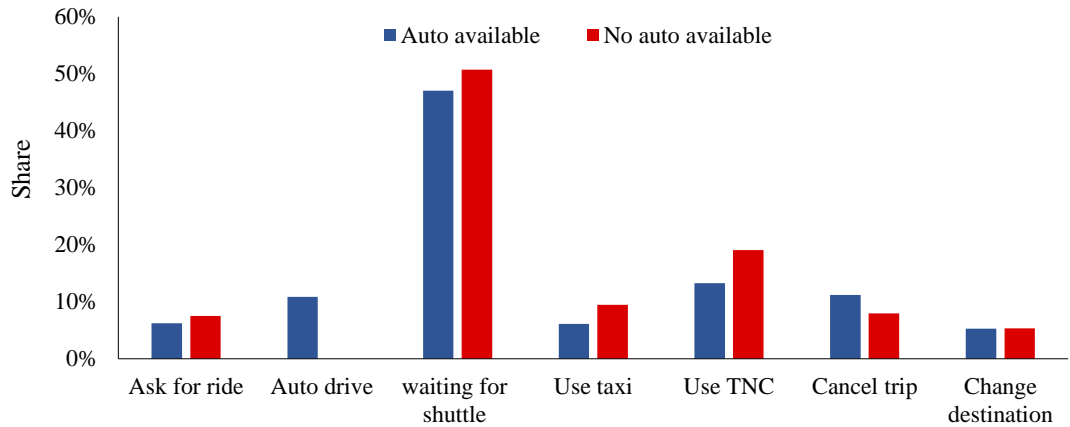

**Figure 2.** The distribution of alternatives in the sample with respect to auto availability

We have also complemented the survey data with the Smart Location Database provided by the Environmental Protection Agency (EPA). The Smart Location Database is a nationwide geographic data which includes population density, land use density, neighborhood design, destination accessibility, transit service, employment, and demographics (Ramsey and Bell, 2014). This information is provided at the census block group level and can provide insights into understanding the effect of built-environment settings on transit users' travel behavior. **Figure 3** depicts the Chicago transit system mapped on the color scheme of two variables: retail employment density and pedestrian-oriented network (that are found to be significant in the final model as will be discussed in the next sections).

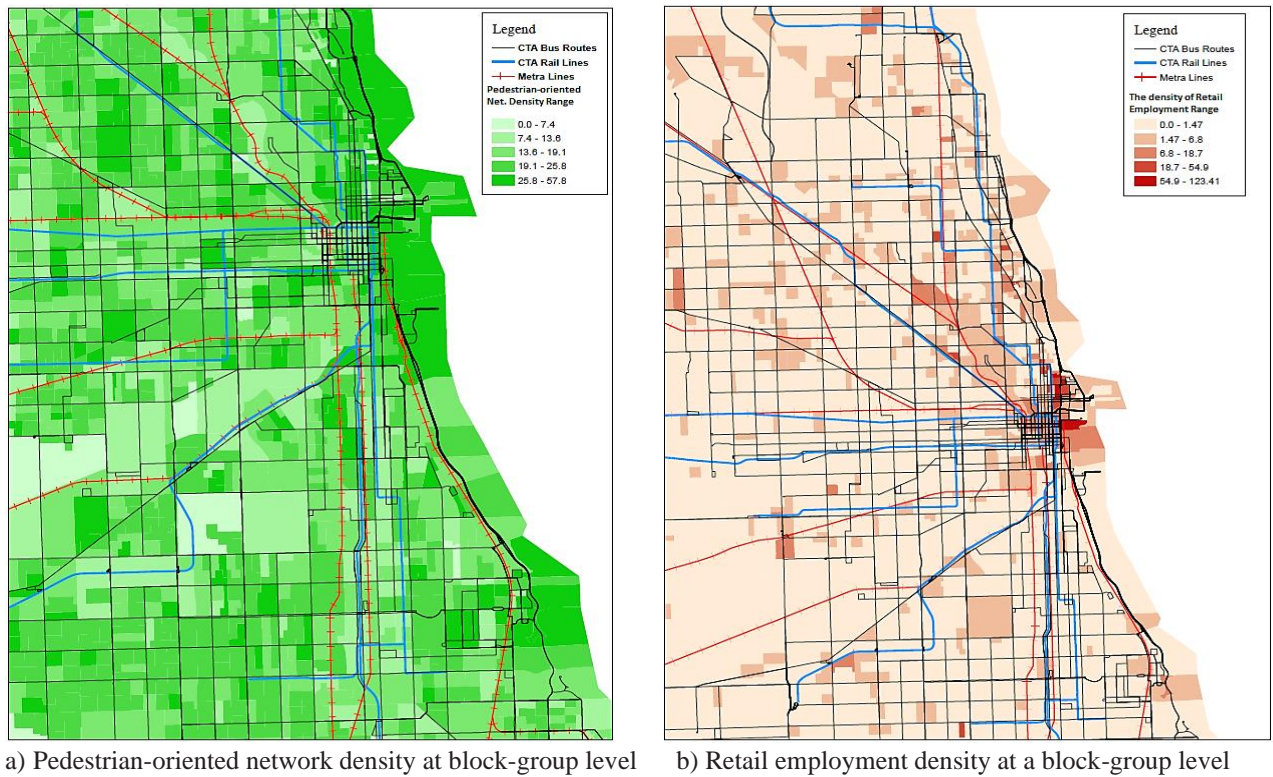

a) Pedestrian-oriented network density at block-group level          b) Retail employment density at a block-group level

**Figure 3.** Chicago transit system at a glance.

Pedestrian-oriented network density represents the level of walkability within a block-group and is calculated by summing pedestrian-oriented links within a block group dividing by the area of that block group (Ramsey and Bell, 2014). The retail employment density for each block group is calculating by summing the total retail jobs within a block-group dividing by the unprotected area of that block-group (Ramsey and Bell, 2014). **Table 2** summarizes the descriptive statistics of the variables used (found to be significant) in the final model.

## METHODOLOGY

In this study, the random parameter multinomial logit (RPMNL) model is applied to understand transit users' decision behavior in case of an unplanned service disruption. This model is highly flexible which obviates the three limitations of multinomial logit (MNL) model by relaxing the independence of irrelevant alternatives (IIA) assumption, allowing for random taste variations, and potential correlation in unobserved factors over time (Greene, 2012; Train, 2009). Consider $U_{int}$ as the utility function of alternative $i$ for decision-maker $n$ in choice situation $t$ as follows:

$$U_{int} = \alpha_{in} + \beta_{in}X_{int} + \varepsilon_{int} \tag{1}$$

where, $\alpha_{in}$ is the constant term for alternative $i$, $\beta_{in}$ represents the estimable coefficients, $X_{int}$ is the vector of explanatory variables of alternative $i$ for decision-maker $n$ in choice situation $t$, and $\varepsilon_{int}$ is the error term. The probability functions of RPMNL model are the integrals of standard logit probabilities over a density of parameters (Train, 2009):

$$P_{nit} = \int \left( \frac{e^{\alpha_{in} + \beta_{in}X_{int}}}{\sum_j e^{\alpha_{jn} + \beta_{jn}X_{jnt}}} \right) f(\beta|\theta) d\beta \tag{2}$$

here, $f(\beta|\theta)$ is the probability density function of coefficients, and $\theta$ would be the parameters that describe the density of $\beta$. Denote $T_n$ as the number of choice situations observed for decision-maker n, then the likelihood function for RPMNL would be (Train, 2009):

$$L_n(\theta) = \prod_{t=1}^{T_n} \prod_{j=1}^{J} \left[ \int \left( \frac{e^{\alpha_{in} + \beta_{in}X_{int}}}{\sum_j e^{\alpha_{jn} + \beta_{jn}X_{jnt}}} \right) f(\beta|\theta) d\beta \right]^{y_{nit}} \tag{3}$$

where, $y_{nit}= 1$ if decision-maker $n$ chooses alternative $i$ in choice situation $t$.

Since, in general, the integrals cannot be solved analytically, the maximum simulated likelihood estimator (MSLE) is suggested to estimate the parameters. In this study, we employed NLOGIT 6.0 to develop the RPMNL model. Also, 200 simulated Halton draws for the model turned to be enough in terms of model stability and accuracy.

**TABLE 2**  Definition Of Variables Used In The Model

| Category | Name | Definition | Mean | Std. Dev. |
|---|---|---|---|---|
| **Demographics** | BACHELOR | 1: If the transit user has a bachelor degree/ 0: Otherwise | 0.386 | 0.487 |
| | GRADUATE | 1: If the transit user has a master's degree and more/ 0: Otherwise | 0.340 | 0.474 |
| | FULL_TIME | 1: If the transit user has a full-time job / 0: Otherwise | 0.723 | 0.447 |
| | MILLENNIAL | 1: If the age of transit user is between 24 and 35/ 0: Otherwise | 0.330 | 0.470 |
| | SENIOR | 1: If the age of transit user is more than 64/ 0: Otherwise | 0.030 | 0.173 |
| | LOW_INCOME | 1: If the household income of transit user is less than $30K | 0.091 | 0.289 |
| **Attitudes** | TRUST | 1: If the transit user trusts and follows the instructions releasing by the transit authority/ 0: Otherwise | 0.765 | 0.423 |
| | RIDESHARE | 1: If the person has the experience of using TNCs (e.g., Uber, Lyft) in the past as a travel mode/ 0: Otherwise | 0.339 | 0.473 |
| | TECH_ACCESS | 1: If the person has access to a smartphone and data/ 0: Otherwise | 0.957 | 0.201 |
| **Trip characteristics** | DISTANCE | The distance between the trip origin and destination in miles | 16.23 | 26.12 |
| | DIST_M15 | 1: If the distance between origin and destination is more than 15 miles / 0: Otherwise | 0.315 | 0.465 |
| | ALONE | 1: If the transit user is traveling alone/ 0: Otherwise | 0.863 | 0.345 |
| | MANDATORY | 1: If the purpose of the trip is work or school/ 0: Otherwise | 0.510 | 0.500 |
| | SHOP | 1: If the trip purpose is shopping/ 0: Otherwise | 0.033 | 0.178 |
| | CTA_RAIL | 1: If the person is waiting for CTA rail/ 0: Otherwise | 0.529 | 0.499 |
| | CTA_METRA | 1: If the person is waiting for Metra rail/ 0: Otherwise | 0.267 | 0.442 |
| | PACE | 1: If the person is waiting for PACE/ 0: Otherwise | 0.042 | 0.201 |
| | SHUTTLE_WAIT | Waiting time for a shuttle bus in minutes | 46.61 | 59.60 |
| | TNC_WAIT | Waiting time for TNC in minutes | 9.55 | 2.84 |
| | TNC_COST | Trip cost for TNC in dollar | 51.53 | 88.85 |
| | DRIVE_TIME | The auto travel time between the trip origin and destination in minutes (Auto, TNC, taxi) | 35.81 | 29.00 |
| | TAXI_WAIT | Waiting time for a taxi in dollar | 21.90 | 15.68 |
| | LONGDIST_MNDT | 1: If the distance between the trip origin and destination is more than 15 miles and the purpose of the trip is work or school./ 0: Otherwise | 0.266 | 0.441 |
| | SHUTTLE _WAIT_METRA | Waiting time for shuttle bus if the disrupted service is Metra transit | 61.56 | 68.47 |
| | SHUTTLE _WAIT_CTA_RAIL | Waiting time for shuttle bus if the disrupted service is CTA rail | 35.87 | 49.18 |
| **Built-environment** | RETAIL_DENSITY | Gross retail employment density in a block group | 3.61 | 9.43 |
| | RET_SHOP | Gross retail employment density in a block group if the trip purpose of transit user is shopping | 0.088 | 0.65 |
| | NDNSTY_PED | Network density regarding facility miles of pedestrian-oriented links per square mile in a block group | 18.98 | 9.46 |
| | NDNSTY_PED_L10 | 1: If NDNSTY_PED < 10/ 0: Otherwise | 0.158 | 0.365 |

## MODEL ESTIMATION & RESULTS

The results of the RPMNL model to analyze transit users' behavior in response to an unplanned service disruption are presented in Table 3. Various variables and variable interactions are tested for each option, and the statistically significant variables at 90%, 95%, and 99% levels of confidence are shown in the table. Based on the results, a wide range of socio-demographic attributes, personal attitudes, trip-related information, and built-environment factors are significant in passengers' response behavior in case of transit service disruptions.

**TABLE 3** Estimation Results Of Random Parameter Multinomial Logit Model

| Explanatory Variable | Coefficient | *p*-value |
|---|---|---|
| CONSTANT (**Auto**) | -0.318 | 0.610 |
|    Std. dev. | 3.035*** | 0.000 |
| CONSTANT (**Shuttle bus**) | 3.210*** | 0.000 |
|    Std. dev. | 2.344*** | 0.000 |
| CONSTANT (**TNC**) | 0.258 | 0.776 |
|    Std. dev. | 1.924*** | 0.000 |
| CONSTANT (**Taxi**) | 0.787** | .0367 |
|    Std. dev. | 1.810*** | 0.000 |
| CONSTANT (**Change destination**) | -3.008*** | 0.000 |
|    Std. dev. | 3.849*** | 0.000 |
| CONSTANT (**Cancel trip**) | -1.003 | 0.172 |
|    Std. dev. | 3.958*** | 0.000 |
| **Ask for ride: base** | | |
| **Auto** | | |
|    LONGDIST_MNDT | 1.594** | 0.045 |
| **Shuttle bus** | | |
|    SHUTTLE_WAIT | -0.015*** | 0.000 |
|    ALONE | 0.220* | 0.091 |
|    SHUTTLE _WAIT_METRA | -0.033*** | 0.000 |
|      Std. dev. | 0.0313*** | 0.000 |
|    SHUTTLE _WAIT_CTA_RAIL | -0.050*** | 0.000 |
|      Std. dev. | 0.082*** | 0.000 |
|    TRUST | 0.607*** | 0.006 |
|    RET_SHOP | -0.324* | 0.070 |
|    NDNSTY_PED_L10 | 0.907** | 0.015 |
|    PACE | 0.587* | 0.089 |
| **TNC** | | |
|    MILLENNIAL | 1.170*** | 0.000 |
|    SENIOR | -2.161** | 0.012 |
|    BACHLOR | 0.608* | 0.073 |
|    GRADUATE | 0.920*** | 0.010 |
|    TNC_WAIT | -0.113*** | 0.002 |
|    TNC_COST | -0.016*** | 0.000 |
|    DRIVE_TIME | -0.017** | 0.040 |
|    TECH_ACCESS | 1.106* | 0.098 |
|    RIDESHARE | 1.019*** | 0.000 |

*** 99% level of confidence, ** 95% level of confidence, * 90% level of confidence

**TABLE 4**  Estimation Results Of Random Parameter Multinomial Logit Model (continued)

| Explanatory Variable | Coefficient | *p*-value |
|---|:---:|:---:|
| **Taxi** | | |
|    SENIOR | -0.887* | 0.1 |
|    FULL_TIME | 0.464* | 0.098 |
|    LOW_INCOME | -0.708* | 0.097 |
|    DRIVE_TIME | -0.023*** | 0.000 |
|    TAXI_WAIT | -0.032*** | 0.000 |
|    RIDESHARE | 0.711* | 0.024 |
| **Change destination** | | |
|    RIDESHARE | -1.596** | 0.030 |
|    Std. dev. | 3.101* | 0.100 |
| **Cancel trip** | | |
|    SENIOR | 0.981* | 0.087 |
|    MANDATORY | -0.751* | 0.091 |
| Number of observations | 2459 | |
| LL (β) | -2733.69 | |
| McFadden Pseudo R-squared | 0.26 | |
| McFadden Pseudo R-squared (multinomial logit model) | 0.07 | |

*** 99% level of confidence, ** 95% level of confidence, * 90% level of confidence

Regarding the auto option, as the results in **Table 3** indicate, long-distance commuters are more likely to use their own vehicle (if accessible) to reach the destination in the case of a transit service disruption. This is possibly because the generalized cost of personal vehicle can be more reasonable compared to TNC or taxi for long-distance travels. This finding is in line with (Limtanakool et al., 2006), who argued that auto is the dominant mode for long-distance commutes. Moreover, due to less flexibility of mandatory trips with respect to the arrival time, choosing personal vehicle can decrease the uncertainty level of arrival time to the planned destination.

With respect to the back-up shuttle bus, it is found that waiting time has a negative impact on the utility of this option. This intuitive finding has been supported by several other studies (e.g., (Chen et al., 2008; Miller et al., 2005) who indicated that higher waiting time can decrease the utility of public transit as travel mode. Moreover, Lin (2017) found that increasing the waiting time for the shuttle bus encourages more transit users to shift to other modes when a service disruption occurs. Our findings, also, add that the influence of waiting time for back-up shuttle bus on passengers' decision is different across various types of transit services.

The variables reflecting the waiting time at the CTA-rail station and the waiting time at the Metra station are both found to be significant with normally distributed random parameters. More specifically, the parameter of waiting time at Metra station and the parameter of waiting time at CTA-rail station are associated with the higher likelihood of switching to other options for the vast majority of observations (i.e., 86% and 73%, respectively). It is possibly because rail users have more concerns about the service to be on-time compared to bus users (Currie and Muir, 2017).

Per the results, Pace users are found to have more tendency to wait for the back-up shuttle bus. As Pace provides bus service to suburban neighborhoods in the Chicago region, people will have less accessibility to other alternatives to reach their destinations or they would be much more expensive compared to Pace service. Thus, Pace users prefer to wait for the back-up shuttle bus in case of an unplanned disruption. Further, as Pace provides special paratransit services for people with physical disabilities, a considerable portion of its users are among such people ("Pace Bus," 2019). Obviously, such people have less flexibility to shift their mode due to their physical constraints.

Also, the indicator parameter of traveling alone is turned to be significant with a positive and fixed parameter across observations. Per the result, transit users who are accompanied by others have less tendency to select the shuttle during a disruption. This is probably because they will have more constraints in scheduling their joint activity-travel compared to an individual traveler. Furthermore, group travelers have the possibility of sharing the cost of the alternative modes which can encourage them to shift to other modes rather than waiting for the recovery shuttle bus.

Our results also indicate that transit users who trust and follow the information provided by transit authorities during a disruption have more tendency to stay at station and wait for the shuttle bus. This finding is in line with the literature arguing that transit users prefer to receive information about the disruption, and this information significantly affects their decision in such situations (Currie and Muir, 2017; Fukasawa et al., 2012; Lin, 2017). However, our study adds that providing information is not expected to have a similar influence on all transit users; rather the influence can be different considering whether passengers trust and follow authorities.

With respect to the built-environment variables, we found that the higher density of retail employment in a block group decreases transit users' likelihood to choose shuttle bus, when their trip purpose is shopping. Also, it is found that within a block group, where the density of pedestrian-oriented links is low, transit users have more tendency to wait for the shuttle bus while a service is disrupted. This result might be because the lower density of pedestrian-oriented links decreases the utility of walking to other destinations or to access other modes (Rodríguez and Joo, 2004), and waiting for shuttle bus could remain the only option with a reasonable disutility.

Turning to the variables that affect the use of ride-sharing services, it is found that millennials have more tendency to choose TNCs while a transit service is disrupted. This finding might be because they are more tech-savvy and familiar with such services compared to older adults (Rayle et al., 2016; Vivoda et al., 2018). In terms of the education level, our results indicate that having a bachelor or graduate degree increases the probability of using ride-sharing services as an alternative to a disrupted transit service. This might be because these people have more knowledge about such services (Vivoda et al., 2018) or they have higher value of time that discourages them to wait for the recovery shuttle bus.

Per the results, the experience of using ride-sharing services in the past can enhance the probability of choosing this mode as an alternative to disrupted transit services. This is possibly because experiencing ride-sharing services, as a relatively new technology in the transportation market, not only could increase the awareness about this mode, but also could improve people's technology acceptance (Wang et al., 2018). Supported by intuition, our results reveal that having access to a smartphone and data positively affects the probability of using TNCs in case of transit disruptions. We also found that, as expected, the trip-related characteristics of TNCs including waiting time, trip cost, and travel time are negatively significant in the model. Mode choice literature supports this finding (e.g. (Chen et al., 2008; Lin, 2017; Miller et al., 2005)) .

With respect to the factors that affect the use of taxi, our results indicate that seniors have less tendency than others to use taxi in case of a transit service disruption. This could be because this group of people have generally more flexibility in scheduling their daily activities and travels, and thus, they might not prefer to choose taxi which is an expensive alternative compared to other modes. This finding is in line with (Rayle et al., 2016) who reported that share of taxi usage among seniors is very low. The results, also, show that low-income passengers are less likely to choose taxi as an alternative to a disrupted transit service. Moreover, having a full-time job is turned to be positively significant in the model. This might be because passengers who have a full-time job have a higher value of time and less flexibility in their activity-travel scheduling.

As expected, our results reveal that waiting time for taxi negatively affects transit users' willing to use this mode during a service disruption. This is in line with (Yang et al., 2000) who argued that passengers consider waiting time for taxi as an important factor of service quality in their decision.

With respect to the variables that affect the decision of changing the destination of the trip, it is found that experiencing a ride-sharing service has a normally-distributed random influence on the probability of this decision. Per the model, the majority of passengers (i.e., about 70% of observations) who experienced ridesharing services in the past have less tendency to change their pre-planned travel destination when a transit service disruption occurs.

As Table 4 indicates, seniors are more likely to cancel their trip when a service disruption occurs. This parameter is found to be fixed across observations. It might be because seniors are most probably retired (Burris and Pendyala, 2002), and have more flexibility in their daily activity-travel scheduling (Frei et al., 2017). We also found that the indicator variable of mandatory trips has a significant and negative effect on the utility of canceling the trip. Supported by intuition, our result indicates that transit riders who are traveling to conduct a mandatory activity (e.g., to workplace or school) have less flexibility to cancel their trip.

**CONCLUSION**

This study presents the results of a new SP-RP survey framework in the Chicago Metropolitan Area about transit users' behavior in response to an unplanned service disruption. In this survey, respondents were intercepted at about 100 separate stops/stations belonging to all major operators of transit service in the region including PACE, Metra, and CTA, and they were given a survey link and unique PIN to access to the questionnaire. In this approach each respondent was faced with four different disruption scenarios in which seven potential actions were listed including waiting for a back-up shuttle bus, asking for a ride from family/friend, picking up his/her own auto (if any), taking a taxi, using ride-sharing services, changing the trip destination, and canceling the trip. Each option, if applicable, was further described in terms of waiting time, travel time, arrival time, and cost. Accounting for heterogeneity across observations as well as panel effects, the random parameter multinomial logit (RPMNL) model is utilized to understand transit users' decision behavior in case of an unplanned service disruption.

According to the results, a wide range of socio-demographic attributes, personal attitudes, trip-related information, and built-environment factors are significant in passengers' response behavior in case of transit service disruptions. Interestingly, our result showed that the effect of service recovery time on passengers are not the same among all types of disrupted services; rail users are more sensitive to the recovery time as compared to bus users. Moreover, although providing information about a service disruption is crucial, our results suggested that the passengers' response to disruption is associated with the fact that how they trust and follow the information. This can provide insights for transit authorities to prepare efficient communication strategies during a service disruption.

The findings of this paper can be used to understand the response of passengers when a service is disrupted. The results also could provide insights for transportation authorities to improve the transit service quality in terms of user's satisfaction and transportation resilience. These insights could help transit agencies in order to implement effective recovery strategies.

## ACKNOWLEDGEMENT

This work was performed under a grant from the Federal Transit Administration awarded to the University of Chicago in December 2016 (IL-26-7015-01 – Coordinated Transit Response Planning). We would like to acknowledge Pace and Metra as representative stakeholders for Chicago's transit operators, providing much of their operating experience and rich data resources to the project. In addition, the CTA has generously provided access to their own data sources and has supported the team by providing access to their facilities for the survey, and has reviewed the survey design during the planning stage. We also thank the Illinois Department of Transportation for their input, as well as other municipal, local, regional, and national stakeholders for their willingness to participate.